# One-dimensional topological quasiperiodic chain for directional wireless power transfer

Juan Song, Fengqing Yang, Zhiwei Guo[*], Youqi Chen, Haitao Jiang, Yunhui Li, and Hong Chen

*Key Laboratory of Advanced Micro-structure Materials, MOE, School of Physics Science and Engineering, Tongji University,*

*Shanghai 200092, China*

**Abstract**

As an important class of systems with unique topological effects beyond the periodic lattices, quasiperiodic topological structures have attracted much attention in recent years. Due to the quasiperiodic modulation, the topological states in the quasiperiodic topological structures have the characteristics of self-similarity, which can be used to observe the charming Hofstadter butterfly. In addition, because of the asymmetric distribution, the edge states in quasiperiodic chain can be used to realize the adiabatic pumping. When the topological parameters in quasiperiodic topological lattices are considered as synthetic dimensions, they can also be used to study the topological properties with higher dimensions. Here, by using ultra-subwavelength resonators, we design and fabricate a type of one-dimensional quasiperiodic Harper chain with asymmetric topological edge states for the directional wireless power transfer (WPT). By further introducing a power source into the system, we selectively light up two Chinese characters which is composed of LED lamps at both ends of the chain. Moreover, the directional WPT implemented by the topological quasiperiodic chain has the property of topological protection, which is immune to the internal disorder perturbation of the structure. Not only do we apply the asymmetric edge state for directional WPT, but also may further actively control the directional WPT by using the external voltage. In addition, this work provides a flexible platform for designing new WPT devices, such as using the corner states in high-order topological structures or the skin effect in the non-Hermitian topological lattices.

**Keywords:** quasiperiodic topological chain; edge states; wireless power transfer;

* Corresponding author: Email: 2014guozhiwei@tongji.edu.cn

**INTRODUCTION**

Recently, a variety of photonic topological models by mapping condensed matter physics have been designed flexibly based on artificial microstructures [1-3]. These photonic topological structures can not only conveniently study various topological phase transitions and the edge states in experiments, but also provide powerful means for electromagnetic wave manipulation [4, 5]. With the rapid development of topological photonics, many physical platforms can be used to study topological physics, such as metamaterials [6-9], coupled resonator array [10-13], photonic crystals [14-16], dielectric resonator chains [17-19], circuit arrays [20-23] and split-ring resonator chains [24-26]. Specially, the topological edge states in classical wave systems can explore many interesting physics which involves the nonlinear [27, 28], non-Hermitian properties [29-37] and quantum optics [38, 39], which enable some unique applications including third-harmonic light [28], sensors [34], lasers [36, 37], and filters [40]. One-dimensional (1D) topological chains have recently been opening exciting directions in the topological photonics. Although the structure of 1D chains is simple, they have rich topological properties, such as topological phase transition, edge inversion, robust edge state and so on. One of the simplest 1D topological chains is the dimer chain, i.e., Su-Schrieffer-Heeger model [41], in which a pair of topological edge sates are localized symmetrically at two ends of the chain [24].

Different from the symmetrical distribution of the edge states in the topological dimer chain, researchers have uncovered the interesting asymmetric edge states in 1D trimer [42]. Quasiperiodic structures are another important class of systems with topological effects [25, 43-48]. Similar to the edge states localized at two ends of the trimer chain, the paired edge states in the quasiperiodic Harper chain are localized at left or right end of the chain [25, 45]. This asymmetric distribution of the edge states may be used in the selective wireless power transfer (WPT) [52, 53]. In 2007, Kurs *et al.* experimentally discovered the magnetic resonance WPT can achieve long-range and high efficient transmission compared with the traditional magnetic induction scheme [54]. Their remarkable work discover the coupling of evanescent waves [55] for WPT and promote the design of new WPT devices [56-58]. Recently, with the development of WPT, more and more functions and application scenarios need to be satisfied. For example, in order to solve the distance dependence of near-field coupling, scientists propose the frequency tracking schemes to realize WPT on move [59-62]. In addition to technical means, the introduction of fascinating non-Hermitian physics [63-65] or metamaterials [66, 67] are also expected to achieve more efficient WPT.

In this work, the asymmetric edge states in a topological quasiperiodic chain are studied for the directional WPT. We theoretically design and experimentally fabricate a finite quasiperiodic Harper chain based on the ultra-subwavelength coil resonators. Based on the near-field measure technology, the density of states (DOS) spectrum of the topological Harper chain is obtained. In addition, the distribution of the edge states in the Harper chain is observed from the local density of states (LDOS) spectrum. By using the asymmetric edge states, we selectively light up two Chinese characters composed of LED lamps at both ends of the chain. Moreover, given the robustness of topological edge state, the designed WPT device will be robust to the disorder perturbation inside the structure. The topological edge states for directional WPT not only extend previous research work on long-range WPT, but they have a circuit structure which is easier to integrate and for active control by considering the variable capacitance diodes into the resonators [68].

**RESULTS**

In the experiment, the 1-D Harper chain is constructed by ultra-subwavelength coil resonator. The coil resonator is placed on a polymethyl methacrylate (PMMA) substrate with thickness of $h = 1$ cm as shown in Fig. 1(a). All the coil resonators in the Harper chain are identical with the resonant frequency of $f_0 = 5.5$ MHz, which is determined by the loaded lumped capacitor $C = 100$ pF and the geometric parameters including the inner diameter $D_1 = 5.2$ cm and outer diameter $D_2 = 7$ cm. In particular, two Chinese characters composed of LED lamps connected to a non-resonant coils are loaded on back layer of the left end and right end of the Harper chain, respectively. The non-resonant coil is shown in Fig. 1(b) and the ports "$A$" and "$B$" are connected with the LED lamps. Once the magnetic field in the top coil is strong enough, the LED lamps loaded in the back layer of the both ends resonators can be lighted up. Fig. 1(c) shows the schematic of a Harper chain with 16 same resonators. In this tight binding model, the quasiperiodic modulation is controlled by tuning the coupling strength.

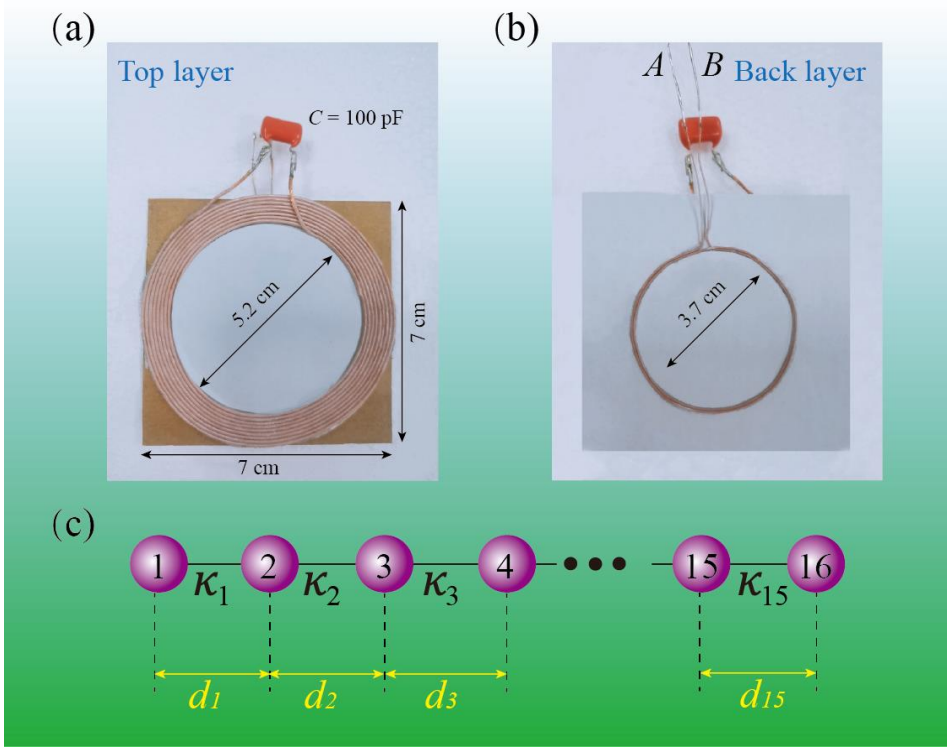

FIG. 1. **Details of the composite resonator and the 1D topological Harper chain.** (a) The resonator on the top layer of the composite structure. Here, the resonant frequency of the coil is 5.5 MHz; the loaded capacitor is 100 pF; the thickness of the substrate is 1 cm; the inner diameter and outer diameter of the coil resonator are 5.2 cm and 7 cm, respectively. (b) A non-resonant coil on the back layer of the two resonators on both ends of the chain. The ports "A" and "B" of the left (right) non-resonant coil are connected with LED lamps. The diameter of the non-resonant coil is 3.7 cm. (c) Schematic of the 1D topological Harper chain. $\kappa_n$ is interpreted as the coupling strength between the $n^{th}$ and $(n+1)^{th}$ resonators, and $d_n$ is the distance, correspondingly.

We start with the Harper model based on the tight-binding mechanism. The topological Harper chain with finite resonators in a quasiperiodic arrangement is controlled by tuning the coupling strength as:

$$\kappa_n = \kappa_0 \left[ 1 - \varepsilon \cos\left(\frac{2\pi}{\tau} n + \phi \right) \right], \tag{1}$$

Where $\kappa_n$ is interpreted as the coupling strength between the $n^{th}$ and $(n+1)^{th}$ resonators, $\kappa_0 = 1$ is a scaling constant, $\varepsilon = 0.5$ is the coefficient that controls the strength of the modulation, and $\tau = (\sqrt{5} + 1)/2$ is the golden ratio. $\phi$ denotes the topological parameter, We vary the value of $\phi$ from 0 to $2\pi$ and obtain the topological band diagram. For the finite-size Harper chain with 16 resonators, we calculate the projected band structure, as shown in Fig. 2(a). The left and right edge states exist in two bandgaps, which are marked by green and red curves, respectively. Here, the value of $\phi$ is 4 marked by black dotted line, to facilitate viewing. The values of the coupling strengths from $\kappa_1$ to $\kappa_{15}$ are calculated from Eq. (1) and shown by the pink dots in Fig. 2(b). Table. 1 shows the corresponding relation between the distances of adjacent resonators and the coupling coefficients.

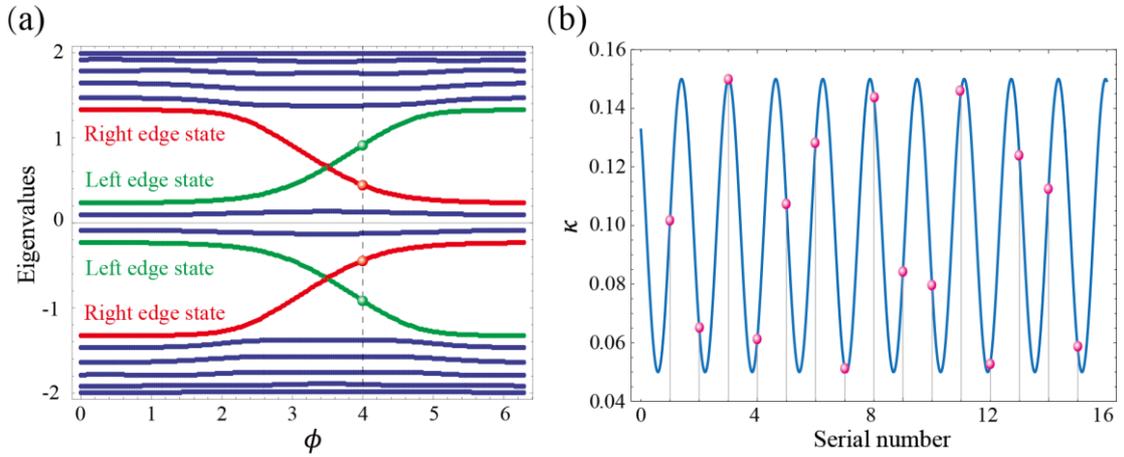

FIG. 2. **Projected band structure and the distribution of coupling coefficients in the Harper chain with 16 resonators**. (a) Projected band structure of the finite-size Harper chain as a function of the topological parameter $\phi$. The left and right edge states exist in the two bandgaps marked by green and red curves, respectively. (b) Distribution of the 15 coupling coefficients based on Eq. (1), in which $\phi = 4$.

Table. 1. Distributions of coupling coefficients and the corresponding separation distances

| Serial number | 1 | 2 | 3 | 4 | 5 | 6 | 7 | 8 |
|---|---|---|---|---|---|---|---|---|
| $\kappa$ | 0.1015 | 0.0652 | 0.1499 | 0.0612 | 0.1073 | 0.1281 | 0.0514 | 0.1437 |
| Distance (cm) | 2.8 | 3.8 | 2 | 3.9 | 2.7 | 2.3 | 4.3 | 2.1 |

| Serial number | 9 | 10 | 11 | 12 | 13 | 14 | 15 |
|---|---|---|---|---|---|---|---|
| $\kappa$ | 0.0842 | 0.0796 | 0.1459 | 0.0527 | 0.1238 | 0.1122 | 0.0583 |
| Distance (cm) | 3.2 | 3.4 | 2 | 4.3 | 2.4 | 2.6 | 4 |

Based on the near-field detection method, we measure the DOS spectrum of the 1D topological Harper chain composed of 16 coil resonators. Our near-field magnetic probe is a loop antenna, which is connected to the port of the vector network analyzer (Agilent PNA Network Analyzer N5222A). The radius of the loop probe is 2 cm. It can be taken as a non-resonant antenna with high impedance. This small loop antenna acts as a source to excite the sample and then measure the reflection. And the LDOS of each site is obtained from the reflection by putting the probe into the center of the corresponding resonator. The DOS spectrum is obtained by averaging the LDOS spectral over all 16 sites. Fig. 3(a) shows the measured DOS spectrum of the topological Harper chain. The left edge state ($f = 5.26$ MHz) and right edge state ($f = 5.45$ MHz) exist in the bandgap of the spectrum. Fig. 3(b) shows the transmission ratio of the two topological edge states, in which the transmission ratio of the left edge state and right edge state ($S_L/S_R$) and the transmission ratio of the right edge state and left edge state ($S_R/S_L$) are marked by green and purple respectively. It can be clearly seen that the $S_L/S_R$ is significantly higher than $S_R/S_L$ at the frequency $f = 5.26$ MHz, and the $S_R/S_L$ is significantly higher than $S_L/S_R$ at the frequency $f = 5.45$ MHz correspondingly, which clearly illustrate that edge states are selectively localized at the left or right ends of the quasiperiodic chain at different frequencies.

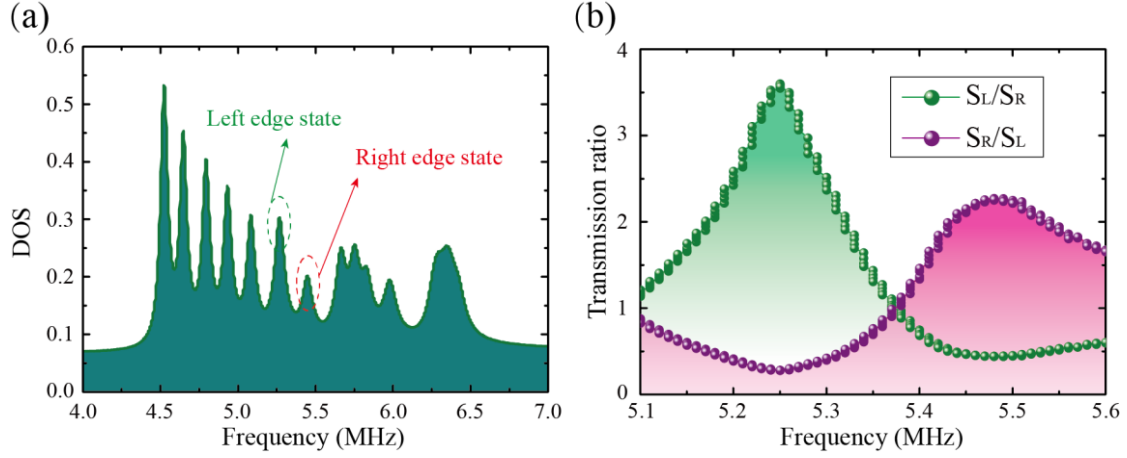

FIG. 3. **Measured DOS spectrum and Transmission ratio of the 1D topological Harper chain**. (a) Measured DOS spectrum of the 1D topological Harper chain. Left edge state ($f$ = 5.26 MHz) and right edge state ($f$ = 5.45 MHz) in the band gap are colored by green arrow and red arrow, respectively. (b) Transmission ratio of the asymmetric topological edge states, the transmission ratio of the left edge state and right edge state ($S_L/S_R$) and the transmission ratio of the right edge state and left edge state ($S_R/S_L$) are marked by green and purple respectively.

Figure 4 shows the LDOS distributions of the left edge state and right edge state, which are marked by the green and red arrows in Fig. 3(a), respectively. For the left edge state, with the frequency $f$ = 5.26 MHz, its LDOS is mainly distributed in the left end of the chain, as shown in Fig. 4(a). And Figure 4(b) shows the LDOS of the right edge state with the frequency $f$ = 5.45 MHz is mainly distributed in the right end of the chain.

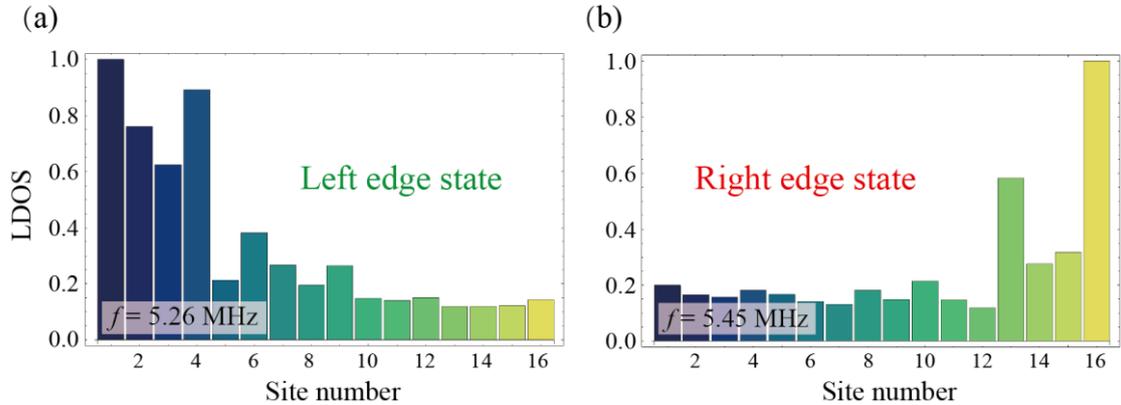

FIG. 4. **Measured normalized LDOS distribution of the left edge state and right edge state**. (a) Measured normalized LDOS distribution of the left edge state (marked by green arrow in Fig. 2(a)) in the 1D topological Harper chain. (b) Measured normalized LDOS distribution of the right edge state (marked by red arrow in Fig. 2(a)) in the 1D topological Harper chain.

The WPT system of magnetic resonance has been demonstrated to be the powerful tools to improve the functionalities and obtain new performance beyond the WPT systems of magnetic induction. In particular, magnetic resonance WPT can achieve long-range and efficient transmission. However, directional wireless power transfer with high efficiency has not been explored. Here, based on the asymmetric topological edge states in the Harper chain, the selective unidirectional wireless power transfer is realized. At the end of this work, the experiment is carried out to exhibit an actually high-power and directional WPT control. The high-power signal source (AG Series Amplifier, T&C Power Conversion) instead of the vector network analyzer, is used to excite the topological edge states. A source non-resonant coil is placed in the center of the structure. To show the topological WPT intuitively, the resonator on left (right) end of the chain is added with the non-resonant coil with the Chinese characters for "tong" ("ji") composed of LED lamps. At the working frequency ($f = 5.26$ MHz) of the left topological edge state, the Chinese character "tong" is lighted up on the left end of the chain whereas the Chinese character "ji" loaded on the right end of the chain remains dark, as shown in Fig. 4(a). When the Harper chain with perturbation, the Chinese character "tong" is still lighted up and Chinese character "ji" is still dark in Fig. 5(b), just as the case without perturbation in Fig. 5(a). However, for the right edge state at the working frequency is 5.45 MHz, whether there is disorder or not, the Chinese character "tong" is always dark on the left end of the chain whereas the Chinese character "ji" loaded on the right end of the chain remains lighted up, as shown in Figs. 5(c) and 5(d). Different from the topological edge states, Figs. 5(e) and 5(f) show that both the Chinese characters "tong" and "ji" are lighted up weakly at the working frequency ($f$ = 5.08 MHz) of the bulk state and then extinguished when perturbation is introduced into the interior of the chain. These results indicate that when perturbation is introduced into the interior of the chain, the topological edge states will not be affected, and the transfer efficiency of the asymmetric edge states is much higher than the transfer efficiency of the bulk state, which are highly significant for robust directional WPT.

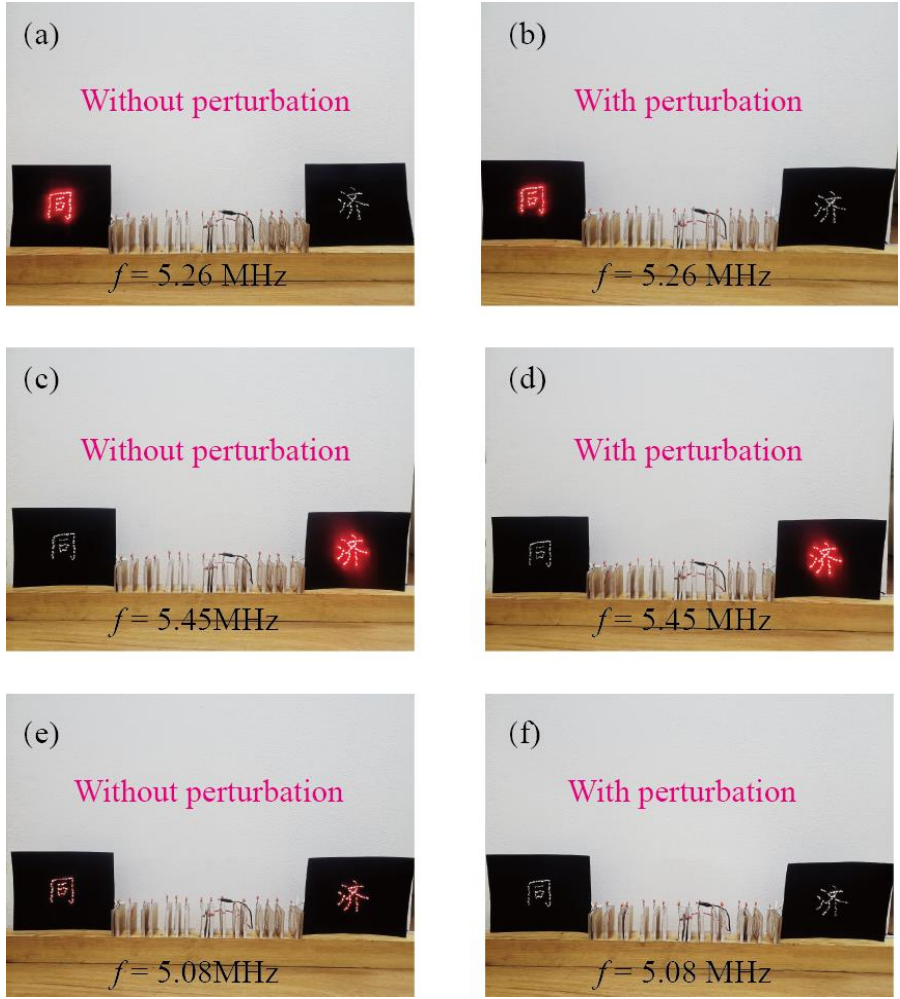

FIG. 5. **Experimental demonstration of the selective directional power transfer for lighting two Chinese characters for "tong" and "ji" with LED lamps.** The non-resonant source coil is placed in the center of the chain. To show the topological WPT intuitively, the resonator on left end of the chain is added with the Chinese characters "tong" and "ji" with LED lamps on left and right end of the chain, respectively. (a) At the frequency $f$ = 5.26 MHz, the left edge state can obviously light up the Chinese characters "tong" on the left end of the chain whereas the Chinese characters "ji" on the right end of the chain remains dark without perturbation. (c) At the frequency $f$ = 5.45 MHz, the right edge state can obviously light up the Chinese characters "ji" on the right end of the chain whereas the Chinese

characters "tong" on the left end of the chain remains dark without perturbation. (e) At the frequency $f =$ 5.08 MHz, the bulk state will slightly light up the Chinese characters "tong" and "ji" without perturbation. (b) (d) (f) Similar to (a) (c) (e), but for the topological Harper chain with perturbation, respectively.

**CONCLUSION**

In summary, based on the 1D topological Harper chain composed of ultra-subwavelength coil resonators, we theoretically and experimentally verify that the asymmetric topological edge states can be used for directional WPT. The asymmetric topological edge states are further used to selectively light up two Chinese characters at two ends of the chain. Specially, this directional WPT is robust to the disorder perturbation inside the structure. The parameters used in this work are all readily accessible for WPT applications. The results for the directional WPT not only provide the directional WPT, but also may facilitate to explore the topological properties in higher-order or more complex setup for WPT, such as the zero-dimensional corner states.

**ACKNOWLEDGMENTS**

This work was supported by the National Key R&D Program of China (Grant No. 2016YFA0301101), the National Natural Science Foundation of China (NSFC; Grant Nos. 11774261, 11474220, and 61621001), the Natural Science Foundation of Shanghai (Grant Nos. 17ZR1443800 and 18JC1410900), the China Postdoctoral Science Foundation (Grant Nos. 2019TQ0232 and 2019M661605), and the Shanghai Super Postdoctoral Incentive Program.